\documentclass[conference]{IEEEtran}
\IEEEoverridecommandlockouts
\usepackage{cite}
\usepackage{amsmath,amssymb,amsfonts}
\usepackage{graphicx}
\usepackage{textcomp}
\usepackage{xcolor}
\usepackage{makecell}
\usepackage{multicol}
\usepackage{siunitx}
\usepackage{booktabs}
\usepackage{xfrac}
\usepackage{algorithm}
\usepackage{algpseudocode}
\usepackage[nolist]{acronym}
\usepackage{tikz}
\usepackage{pgfplots}
\usepackage{rotating}
\pgfplotsset{width=7cm,compat=1.3}
\usetikzlibrary{shapes,arrows,positioning,calc, matrix,spy,patterns}
\usetikzlibrary{decorations.pathreplacing,calligraphy}
\usetikzlibrary{colorbrewer}
\usetikzlibrary{fit}
\usetikzlibrary{external}
\usepgfplotslibrary{colorbrewer}

\definecolor{cb-1}{HTML}{4477AA}
\definecolor{cb-2}{HTML}{EE6677}
\definecolor{cb-3}{HTML}{228833}
\definecolor{cb-4}{HTML}{CCBB44}
\definecolor{cb-5}{HTML}{66CCEE}
\definecolor{cb-6}{HTML}{AA3377}
\definecolor{cb-7}{HTML}{BBBBBB}

\pgfplotscreateplotcyclelist{cb list}{
	cb-1,cb-2,cb-3,cb-4,cb-5,cb-6,cb-7
}

\definecolor{kit-green100}{rgb}{0,.59,.51}
\definecolor{kit-green70}{rgb}{.3,.71,.65}
\definecolor{kit-green50}{rgb}{.50,.79,.75}
\definecolor{kit-green30}{rgb}{.69,.87,.85}
\definecolor{kit-green15}{rgb}{.85,.93,.93}
\definecolor{KITgreen}{rgb}{0,.59,.51}

\definecolor{KITpalegreen}{RGB}{130,190,60}
\colorlet{kit-maigreen100}{KITpalegreen}
\colorlet{kit-maigreen70}{KITpalegreen!70}
\colorlet{kit-maigreen50}{KITpalegreen!50}
\colorlet{kit-maigreen30}{KITpalegreen!30}
\colorlet{kit-maigreen15}{KITpalegreen!15}

\definecolor{KITblue}{rgb}{.27,.39,.66}
\definecolor{kit-blue100}{rgb}{.27,.39,.67}
\definecolor{kit-blue70}{rgb}{.49,.57,.76}
\definecolor{kit-blue50}{rgb}{.64,.69,.83}
\definecolor{kit-blue30}{rgb}{.78,.82,.9}
\definecolor{kit-blue15}{rgb}{.89,.91,.95}

\definecolor{KITyellow}{rgb}{.98,.89,0}
\definecolor{kit-yellow100}{cmyk}{0,.05,1,0}
\definecolor{kit-yellow70}{cmyk}{0,.035,.7,0}
\definecolor{kit-yellow50}{cmyk}{0,.025,.5,0}
\definecolor{kit-yellow30}{cmyk}{0,.015,.3,0}
\definecolor{kit-yellow15}{cmyk}{0,.0075,.15,0}

\definecolor{KITorange}{rgb}{.87,.60,.10}
\definecolor{kit-orange100}{cmyk}{0,.45,1,0}
\definecolor{kit-orange70}{cmyk}{0,.315,.7,0}
\definecolor{kit-orange50}{cmyk}{0,.225,.5,0}
\definecolor{kit-orange30}{cmyk}{0,.135,.3,0}
\definecolor{kit-orange15}{cmyk}{0,.0675,.15,0}

\definecolor{KITred}{rgb}{.63,.13,.13}
\definecolor{kit-red100}{cmyk}{.25,1,1,0}
\definecolor{kit-red70}{cmyk}{.175,.7,.7,0}
\definecolor{kit-red50}{cmyk}{.125,.5,.5,0}
\definecolor{kit-red30}{cmyk}{.075,.3,.3,0}
\definecolor{kit-red15}{cmyk}{.0375,.15,.15,0}

\definecolor{KITpurple}{RGB}{160,0,120}
\colorlet{kit-purple100}{KITpurple}
\colorlet{kit-purple70}{KITpurple!70}
\colorlet{kit-purple50}{KITpurple!50}
\colorlet{kit-purple30}{KITpurple!30}
\colorlet{kit-purple15}{KITpurple!15}

\definecolor{KITcyanblue}{RGB}{80,170,230}
\colorlet{kit-cyanblue100}{KITcyanblue}
\colorlet{kit-cyanblue70}{KITcyanblue!70}
\colorlet{kit-cyanblue50}{KITcyanblue!50}
\colorlet{kit-cyanblue30}{KITcyanblue!30}
\colorlet{kit-cyanblue15}{KITcyanblue!15}

\definecolor{KITbraun}{RGB}{167,130,46}

\definecolor{cb-1}{HTML}{4477AA}
\definecolor{cb-2}{HTML}{EE6677}
\definecolor{cb-3}{HTML}{228833}
\definecolor{cb-4}{HTML}{CCBB44}
\definecolor{cb-5}{HTML}{66CCEE}
\definecolor{cb-6}{HTML}{AA3377}
\definecolor{cb-7}{HTML}{BBBBBB}

\pgfplotscreateplotcyclelist{cb list}{
	cb-1,cb-2,cb-3,cb-4,cb-5,cb-6,cb-7
}

\let\Re\relax

\DeclareMathOperator{\Re}{Re}

\newcommand{\RV}[1]{\mathsf{#1}}

\newcommand*{\vect}[1]{\boldsymbol{#1}}

\newcommand*{\e}{\mathrm{e}}
\let\j\relax
\newcommand{\j}{\mathrm{j}}

\newcommand{\expecv}[2]{\mathbb{E}_{#1} \! \left\{#2\right\}}

\newcommand*{\PD}{P_{\text{D}}}
\newcommand*{\PFA}{P_{\text{FA}}}

\newcommand*{\hHn}{\hat{\RV{H}}_n}

\usepackage{xcolor}
\usepackage{amsmath}
\usepackage{amssymb}
\usetikzlibrary{calc}
\usetikzlibrary{positioning}
\usetikzlibrary{patterns}
\usetikzlibrary{decorations.pathreplacing,calligraphy}

\tikzset{
	lsblock/.style={rectangle, thick, draw, minimum width=1.2cm, minimum height=0.6cm, rounded corners=1.6mm, font=\small, align=center}
}

\tikzset{
	aeblock/.style={rectangle, thick, draw, minimum width=2.4cm, minimum height=0.7cm, rounded corners=1.6mm, align=center}
}
\tikzset{
	smlblock/.style={rectangle, thick, draw, minimum width=1.8cm, minimum height=0.6cm, rounded corners=1.2mm, align=center}
}
\tikzset{
	smsblock/.style={rectangle, thick, draw, minimum width=0.6cm, minimum height=0.4cm, rounded corners=1.0mm,inner sep=0.05cm, align=center}
}

\tikzstyle{surround} = [rectangle, rounded corners, draw=KITred, inner sep=0.25cm, dashed, thick]
\tikzstyle{surround_AE} = [rectangle, rounded corners, draw=KITred, inner sep=0.13cm, dashed, thick]

\def\BibTeX{{\rm B\kern-.05em{\sc i\kern-.025em b}\kern-.08em
    T\kern-.1667em\lower.7ex\hbox{E}\kern-.125emX}}

\begin{document}
\begin{acronym}[]
    \acro{6G}{sixth generation}
    \acro{AE}{autoencoder}
    \acro{AF}{ambiguity function}
    \acro{AIR}{achievable information rate}
    \acro{AWGN}{additive white Gaussian noise}
    \acro{CA}{cell-averaging}
    \acro{CFAR}{constant false alarm rate}
    \acro{CP}{cyclic prefix}
    \acro{FEC}{forward error correction}
    \acro{FFT}{fast Fourier transform}
    \acro{GMI}{generalized mutual information}
    \acro{IFFT}{inverse fast Fourier transform}
    \acro{ISAC}{integrated sensing and communications}
    \acro{LLR}{log-likelihood ratio}
    \acro{MF}{matched filter}
    \acro{OFDM}{orthogonal frequency division multiplexing}
    \acro{PSK}{phase shift keying}
    \acro{RCS}{radar cross section}
    \acro{QAM}{quadrature amplitude modulation}
    \acro{RV}{random variable}
    \acro{SC}[S\&C]{sensing \& communications}
    \acro{SD}{soft decision}
    \acro{SINR}{signal-to-interference-and-noise ratio}
    \acro{SNR}{signal-to-noise ratio}
    \acro{TOI}{target of interest}
\end{acronym}

\title{Joint Optimization of Geometric and Probabilistic Constellation Shaping for OFDM-ISAC Systems
\thanks{This work has received funding from the German Federal Ministry of Education and Research (BMBF) within the projects Open6GHub (grant agreement 16KISK010) and KOMSENS-6G (grant agreement 16KISK123). Mr. Geiger acknowledges the Networking Grant from the Karlsruhe House of Young Scientists.}
}

\author{
\IEEEauthorblockN{Benedikt Geiger\IEEEauthorrefmark{1}, Fan Liu\IEEEauthorrefmark{2}, Shihang Lu\IEEEauthorrefmark{2}, Andrej Rode\IEEEauthorrefmark{1}, Laurent Schmalen\IEEEauthorrefmark{1}}
\IEEEauthorblockA{\IEEEauthorrefmark{1}\ Communications Engineering Lab (CEL), Karlsruhe Institute of Technology (KIT), 76187 Karlsruhe, Germany\\
\IEEEauthorrefmark{2}\ Southern University of Science and Technology, China\\
Email: \texttt{benedikt.geiger@kit.edu}}
}

\maketitle

\begin{abstract}
6G communications systems are expected to integrate radar-like sensing capabilities enabling novel use cases. However, \ac{ISAC} introduces a trade-off between communications and sensing performance because the optimal constellations for each task differ. In this paper, we compare geometric, probabilistic and joint constellation shaping for \ac{OFDM}-\ac{ISAC} systems using an \ac{AE} framework. We first derive the constellation-dependent detection probability and propose a novel loss function to include the sensing performance in the \ac{AE} framework. Our simulation results demonstrate that constellation shaping enables a dynamic trade-off between communications and sensing. Depending on whether sensing or communications performance is prioritized, geometric or probabilistic constellation shaping is preferred. Joint constellation shaping combines the advantages of geometric and probabilistic shaping, significantly outperforming legacy modulation formats.
\end{abstract}

\acresetall
\section{Introduction}
\Ac{ISAC} is expected to equip communications networks with a sixth sense by utilizing the \ac{OFDM} communication signals for radar-like sensing. This joint design reduces hardware complexity, improves energy efficiency, and enhances reliability compared to operating two separate systems~\cite{wild_6g_2023,liu_integrated_2022,Lu_challenges}.

However, the choice of the modulation constellation impacts \ac{SC} performance resulting in the random-deterministic trade-off~\cite{keskin_fundamental_2024,xiong_torch_2024}. In essence, a Gaussian distributed constellation maximizes the communications performance for an \ac{AWGN} channel, whereas constant modulus constellations maximize sensing performance~\cite{xiong_torch_2024}. 
These contradicting requirements on the constellation make constellation shaping a crucial tool to balance the \ac{SC} performance in practical \ac{ISAC} systems.

There are three main approaches to constellation shaping: \mbox{$i)$
\textit{Geometric shaping:}} optimizes the location of constellation points, assuming an equal probability of occurrence for each point. \mbox{$ii)$ \textit{Probabilistic shaping:}} uses conventional \ac{QAM} constellation points but optimizes the probability distribution of these points. \mbox{$iii)$ \textit{Joint shaping:}} optimizes both location and probability of the constellation points. 
These different methods have been successfully optimized and compared for communications using an \ac{AE} framework~\cite{stark_joint_2019,aref_end--end_2022}. In these works, the communications system is modeled by differentiable blocks and the constellation points and their probabilities are treated as trainable parameters.

Recently, probabilistic shaping was investigated for \ac{ISAC} to trade off mutual information against side-lobe level in the \ac{AF} to balance \ac{SC} performance~\cite{du_reshaping_2023, Yang_Constellation_Design}. However, the \ac{AF} does not capture the detection performance in practical \ac{ISAC} systems in an end-to-end manner. This raises two fundamental questions: How does the constellation affect the detection probability in practical \ac{OFDM}-\ac{ISAC} systems? How do different approaches to constellation shaping influence the \ac{ISAC} trade-off?

In this paper, we address these research gaps by first deriving the impact of the constellation on the detection probability, and employ different constellation shaping methods to balance \ac{SC} performance in an end-to-end manner. Second, we employ a bitwise \ac{AE} to optimize geometric, probabilistic, and joint constellations, maximizing the \ac{GMI} under specific detection and false alarm constraints. Finally, we validate our analytical results through simulations and compare the different approaches demonstrating that our proposed joint optimization outperforms both geometric and probabilistic constellation shaping.

\begin{figure*}[!h]
    \centering
    \input{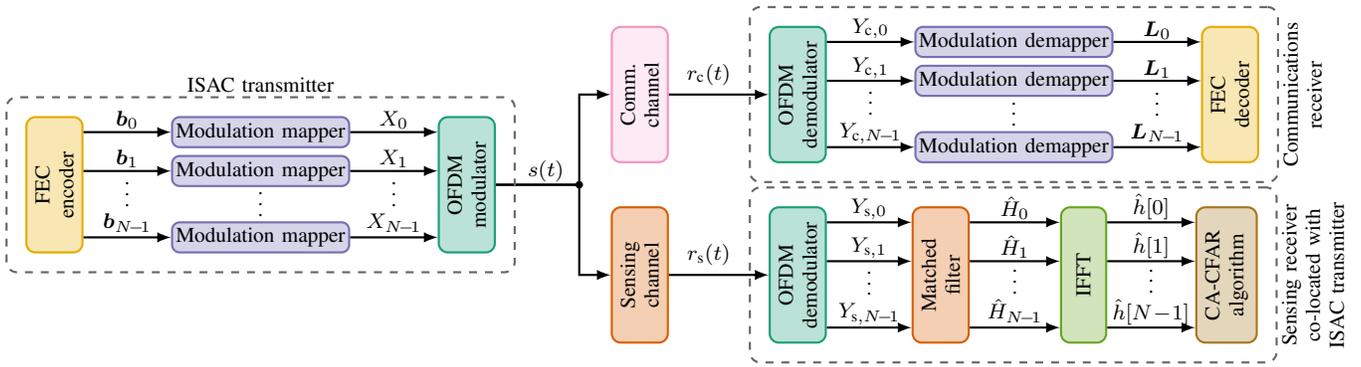}
    \vspace{-0.30cm}
    \caption{Block diagram of the considered monostatic bit-interleaved coded modulation \ac{OFDM}-\ac{ISAC} system}
    \label{fig:system_model:block_diagram}
    \vspace{-0.3cm}
\end{figure*}

\section{System model}
\label{sec:ch2:system_model}
In this work, we consider a monostatic bit-interleaved coded modulation \ac{OFDM}-\ac{ISAC} system as illustrated in Fig.~\ref{fig:system_model:block_diagram}. To reduce computational complexity and simplify the analysis, we assume that all sub-carriers apply the same constellation, and the delays of the targets are assumed to be constant multiples of the sampling time shorter than the duration of the \ac{CP}. Given that the targets are static, we focus on the transmission of single \ac{OFDM} symbols and neglect velocity (Doppler) estimation.%

\subsection{\acs{ISAC} Transmitter}
For each sub-carrier \mbox{$n \in \{0, \ldots, N \! - \! 1\}$} of the \ac{OFDM} system, a constellation mapper encodes $M$ bits \mbox{$\vect{b}_n \in \{0,1\}^{M}$}, obtained from a binary \ac{FEC} encoder, onto one of \mbox{$\tilde{M} = 2^M$} modulation symbols $X$ in the modulation alphabet $\mathcal{X}$, i.e., \mbox{$X \in \mathcal{X} \subset \mathbb{C}$}, where $\left| \mathcal{X} \right| = \tilde{M}$. Since the constellation symbols are randomly selected, the transmit symbols $\RV{X}_n$ can be considered as i.i.d. \acp{RV} \mbox{$\RV{X}_n \sim P(X)$} with \acp{RV} being denoted by sans-serif font.

The \ac{OFDM} modulator then transforms the $N$ frequency domain symbols $\RV{X}_n$, into the time domain using the orthonormal \ac{IFFT}. Next, a \ac{CP} is added before the baseband signal $\RV{s}(t)$ is upconverted to the carrier frequency and transmitted.

\subsection{Sensing Channel and Receiver}
The arbitrary but fixed sensing channel $h(t)$ is modeled assuming $J$ static point targets, each with a delay $\tau_j$ and a complex amplitude $a_j$. The received sensing signal is
\begin{equation}
    \RV{r}_{\text{s}}(t) = h(t) * \RV{s}(t) + \RV{w}_{\text{s}}(t) = \sum_{j=1}^{J} a_j \RV{s}(t - \tau_j) + \RV{w}_{\text{s}}(t), 
\end{equation}
where $\RV{w}_{\text{s}}(t) \! \sim \! \mathcal{CN} \left( 0, \sigma_{\text{s}}^2 \right)$ denotes \ac{AWGN} with variance $\sigma_{\text{s}}^2$. 
The sensing receiver samples the baseband signal $\RV{r}_{\text{s}}(t)$, removes the \ac{CP}, and transforms it into the frequency domain using the orthonormal \ac{FFT}. The frequency domain received symbols can then be expressed as
\vspace{-0.3cm}
\begin{equation}
    \RV{Y}_{\text{s},n} = \RV{X}_n H_n + \RV{W}_{\text{s},n} = \RV{X}_n \frac{1}{\sqrt{N}} \sum_{j=1}^{J} a_j \e ^{- \j 2 \pi \frac{n}{N} \tau_j} + \RV{W}_{\text{s},n},
\end{equation}
where $H_n$ denotes the channel transfer function and $\RV{W}_{\text{s},n}$ is the \ac{FFT} of the sampled \ac{AWGN} $\RV{w}_{\text{s}}(t)$, which follows $\mathcal{CN} \left( 0, \sigma_{\text{s}}^2 \right)$.

The sensing receiver applies a sensing \ac{MF} 
\begin{equation}
    \hat{\RV{H}}_n = \RV{Y}_{\text{s},n} \RV{X}^{*}_n = (H_n \RV{X}_n \! + \! \RV{W}_{\text{s},n} ) \RV{X}^{*}_n = H_n \left| \RV{X}_n \right| ^2 \! + \! \RV{W}_{\text{s},n} \RV{X}^{*}_n,  %
    \label{eq:system_model:transfer_function_estimate}
\end{equation}
which is an unbiased estimate of the sensing channel for unit power constellations $\mathbb{E}_{\RV{X}} \{ | \RV{X}_n  |^2  \} = \num{1}$
\begin{equation}
    \mathbb{E}_{\RV{X}} \{\hat{\RV{H}}_n\} = \mathbb{E}_{\RV{X}} \{ H_n | \RV{X}_n  |^2  \} + \expecv{\RV{X}}{ \RV{W}_{\text{s},n} \RV{X}^{*}_n } = H_n.
    \label{eq:system_model:mean}
\end{equation}

The delay domain channel estimate $\hat{\RV{h}}[k]$ is obtained by applying the orthonormal \ac{IFFT} to the frequency domain channel estimate $\hat{\RV{H}}_n$
\begin{equation}
    \hat{\RV{h}}[k]= \frac{1}{\sqrt{N}} \sum_{n=0}^{N-1} \hat{\RV{H}}_n \mathrm{e}^{\mathrm{j} 2 \pi \frac{n}{N} k}, \qquad k = 0, \ldots N-1.
\end{equation}

Finally, the \ac{CA}-\ac{CFAR} algorithm, which maximizes the detection probability $\PD$ given a maximum false alarm rate $\PFA$, determines whether a target is present at a delay $k$~\cite{richards_fundamentals_2014}.

\subsection{Communications Channel and Receiver}
For communications, the influence of a potential multi-path communications channel is assumed to be suppressed by classical equalization methods and the channel is modeled as an \ac{AWGN} channel with noise variance $\sigma^2_{\text{c}}$, represented by \mbox{$\RV{r}_{\text{c}}(t) = \RV{s}(t) + \RV{w}_{\text{c}}(t)$}, where \mbox{$\RV{w}_{\text{c}}(t) \sim \mathcal{CN} \left( 0, \sigma_{\text{c}}^2 \right)$}. Similarly to the sensing receiver, the communications receiver removes the \ac{CP} from the baseband signal $\RV{r}_{\text{c}}(t)$ and converts it into the frequency domain using an orthonormal \ac{IFFT}. This simplifies the \ac{OFDM} system into a set of $N$ parallel \ac{AWGN} channels, one for each sub-carrier. The received communications symbol of the $n$th sub-carrier after equalization can therefore be modeled as
\vspace{-0.2cm}
\begin{equation}
    \RV{Y}_{\text{c},n} = \RV{X}_n + \RV{W}_{\text{c},n},
    \label{eq:system_model:comm_channel}
\end{equation}
where $\RV{W}_{\text{c},n} \sim \mathcal{CN}(0, \sigma_{\text{c},n}^2)$ is \ac{AWGN} with variance $\sigma_{\text{c},n}^2$. 

The constellation demapper computes the \acp{LLR} for each bit~\cite{Ivanov_BICM}
\begin{equation}
    \RV{L}_{n,m} \! = \!  \log \frac{\sum_{X_{\!n} \in \mathcal{X}^{0}_m}{ \! f_{\RV{Y}_{\text{c},n}|\RV{X}_n}(Y_{\text{c},n}|X_n)}}{\sum_{X_n \in \mathcal{X}^{1}_m}{ \! f_{\RV{Y}_{\text{c},n}|\RV{X}_n}(Y_{\text{c},n}|X_n)}}, \, m \! = \! 1, \ldots, M,
    \label{eq:system_model:LLR}
\end{equation}
where $\mathcal{X}^{b}_m$ represents the set of constellation symbols labeled with bit $b \in \{0,1\}$ at bit position $m$ and $f_{\RV{Y}_{\text{c},n}|\RV{X}_n}$ denotes the communications channel transition probability density function. The \acp{LLR} are a measure of the reliability of each bit, where a large magnitude indicates high reliability and a small magnitude indicates low reliability. These \acp{LLR} are eventually fed to a \ac{SD}-\ac{FEC} decoder.

To evaluate the communications performance, we use the \ac{GMI}, which is a performance measure for the \ac{AIR} in practical bit-interleaved coded modulation systems\cite{Ivanov_BICM}
\begin{equation}
    \mathrm{GMI}_n = \sum_{m=1}^{M} I(\RV{b}_{n,m};\RV{L}_{n,m}), %
    \label{eq:system_model:GMI}
\end{equation}
where $I(\RV{b}_{n,m};\RV{L}_{n,m})$ denotes the bitwise mutual information between the $m$th bit and corresponding \ac{LLR} of the $n$th sub-carrier. 

\subsection{The Constellation-dependent Detection Probability}
Since estimating the detection probability $P_{\mathrm{D}}$ using Monte Carlo simulations results in high computational complexity, we derive the constellation-dependent detection probability by analyzing each signal processing block of the sensing receiver. For the \ac{CA}-\ac{CFAR} algorithm, the detection probability
\begin{equation}
    \PD =  \PFA ^{\frac{1}{1 + \gamma}},
    \label{eq:system_model:detection_prob}
\end{equation}
assuming Gaussian distributed noise and interference~\cite{richards_fundamentals_2014} depends only on the false alarm rate $\PFA$ and the average \ac{SINR} $\gamma$ of the signal fed to the detection algorithm. Since the false alarm rate $\PFA$ is usually fixed during system design, increasing the detection probability $\PD$ requires increasing the average \ac{SINR} $\gamma$ at the input of the \ac{CA}-\ac{CFAR}, i.e., at the output of the \ac{IFFT}.

\begin{figure}[!b]
    \vspace{-0.5cm}
	\centering
	\input{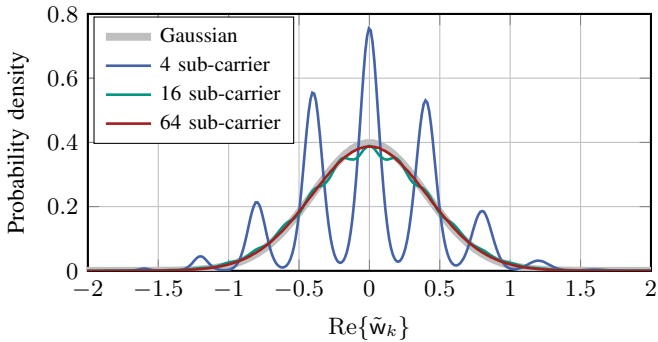}
    \vspace{-0.8cm}
	\caption{Probability distribution of the noise at the input of the target detector, i.e., at the output of the \ac{IFFT} assuming a \num{16}-\ac{QAM} and a sensing \ac{SNR} of \SI{20}{dB} in a single target scenario for various numbers of sub-carriers.}
	\label{fig:system_model:noise_PDF}
\end{figure}

However, when random constellation symbols are transmitted, the noise at the output of the \ac{MF}, i.e., at the input of the \ac{IFFT}, may no longer be Gaussian as can be observed from~(\ref{eq:system_model:transfer_function_estimate}). To show that the noise is still approximately Gaussian at the output of the \ac{IFFT}, we decompose the channel transfer function estimate $\hat{\RV{H}}_n$ into a deterministic part $H_n$ and a random part $\tilde{\RV{W}}_n$, which accounts for \ac{AWGN} and the randomness of the modulation, i.e. \mbox{$\hat{\RV{H}}_n = H_n + \tilde{\RV{W}}_n$}.

For the random part $\tilde{\RV{W}}_n$, the \ac{IFFT} acts as a summation of $N$ independent and scaled \acp{RV} $\tilde{\RV{W}}_n$. According to the central limit theorem, the random part of the channel estimate \mbox{$\tilde{\RV{w}}[k] = \hat{\RV{h}}[k] - h[k]$}, which is the \ac{IFFT} of $\tilde{\RV{W}}_{n}$ approximates a Gaussian if the number of sub-carriers $N$ is sufficiently large. Fig.~\ref{fig:system_model:noise_PDF} shows the probability distribution of $\Re \{\tilde{\RV{w}}[k]\}$ for an increasing number of sub-carriers, demonstrating that already \num{64} sub-carriers are sufficient for the Gaussian assumption. This justifies modeling the detection probability using~(\ref{eq:system_model:detection_prob}).

The variance 
\begin{equation}
\sigma^2_{\hat{\RV{h}}[k]} = \left( \frac{1}{\sqrt{N}} \right)^{\!2} \sum_{n = 0}^{N-1} \sigma^2_{\hHn} = \frac{1}{N} N \sigma^2_{\hHn} = \sigma^2_{\hHn}
\end{equation}
of the delay domain channel estimate equals the variance of the frequency domain channel estimate. For the deterministic part of the channel estimate, the \ac{IFFT} leads to an integration gain, increasing the power of the targets and consequently the \ac{SINR} by a factor of $N$, as
\begin{equation}
    \frac{1}{\sqrt{N}} \sum_{n=0}^{N-1} \sum_{j=1}^{J} a_j \mathrm{e}^{-\mathrm{j} 2 \pi \frac{n \tau_j}{N}} \mathrm{e}^{\mathrm{j} 2 \pi \frac{n k}{N}} = \begin{cases}
        \sqrt{N} a_j, & k \! = \! \tau_j \\
        0 & k \! \neq \! \tau_j.
    \end{cases}
    \label{eq:system_model:IFFT_deterministic}
\end{equation}
Therefore, the average \ac{SINR} and consequently the detection probability $P_{\mathrm{D}}$ is increased by reducing the variance
\begin{align}
        \sigma^2_{\hHn} & = \mathbb{E}_{\RV{X}} \{ | \hHn |^2  \} - | \mathbb{E}_{\RV{X}} \{\hHn\} | ^2 \nonumber \\
 & = | H_n |^2 ( \mathbb{E}_{\RV{X}} \{ \left| \RV{X} \right|^4 \} - 1 )  +\sigma_{\mathrm{s}}^2 \label{eq:system_model:variance_freq_domain} \\
 & = | H_n |^2 \left( \kappa - 1 \right)  +\sigma_{\mathrm{s}}^2, \nonumber
\end{align}
where $\kappa$ is the kurtosis of the constellation, which is equivalent to the 4th-order moment for unit power zero mean constellations. Note that unit modulus constellations like \ac{PSK} have the lowest kurtosis of \mbox{$\kappa = \num{1}$}, leading to the highest detection probability. A circular complex Gaussian distribution, which maximizes the \ac{AIR} for an \ac{AWGN} channel, has a kurtosis of \mbox{$\kappa = \num{2}$} and results in a lower detection probability~\cite{liu_ofdm_2024}. From~(\ref{eq:system_model:IFFT_deterministic}) and~(\ref{eq:system_model:variance_freq_domain}) follows the average \ac{SINR}
\vspace{-0.3cm}
\begin{equation}
    \gamma_{\mathrm{TOI}} = \frac{N \cdot | a_{\mathrm{TOI}} |^2}{ \sum_{j=1}^{J} |a_j |^2 ( \kappa -1 )  + \sigma_{\text{s}}^2 }.
    \label{eq:system_model:SNR_TOI}
\end{equation}
of a \ac{TOI} with complex amplitude $a_{\mathrm{TOI}}$ at the input of the \ac{CFAR}. Inserting~(\ref{eq:system_model:SNR_TOI}) into (\ref{eq:system_model:detection_prob}) yields the constellation-dependent detection probability. We observe that the detection probability $\PD$ depends only on the kurtosis $\kappa$ of the constellation for a given scenario.

\begin{figure*}[!h]
    \centering
    \input{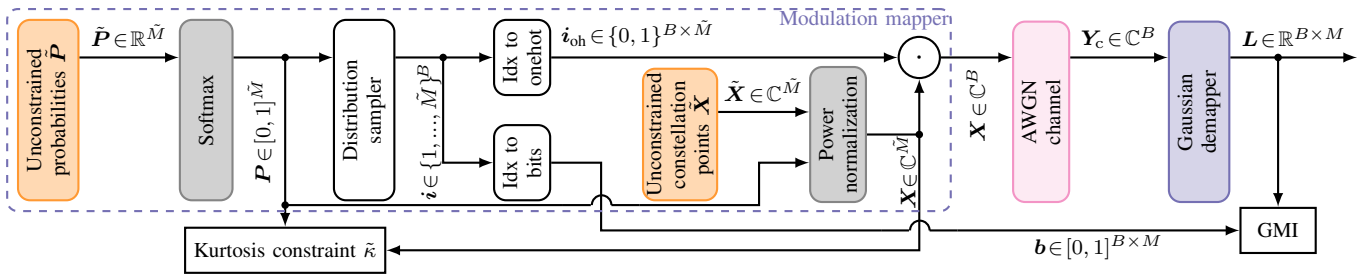}
    \vspace{-0.7cm}
    \caption{Proposed \ac{AE} framework to shape constellations for \ac{ISAC}. The trainable parameters are marked in orange and their normalization in gray.}
    \label{fig:AE:setup}
    \vspace{-0.3cm}
\end{figure*}

\subsection{Optimization Problem}
We aim to find the constellation that maximizes the \ac{GMI} of the overall \ac{ISAC} system, subject to a minimum detection probability constraint $\alpha_{\mathrm{D}}$, i.e. \mbox{$P_{\mathrm{D}} \geqslant \alpha_{\mathrm{D}}$}. Since all sub-carriers use the same constellation, the total \ac{GMI} is maximized if the \ac{GMI} per sub-carrier is maximized. Further, we reformulate the detection probability constraint $\alpha_{\mathrm{D}}$ as a kurtosis constraint $\tilde{\kappa}$, i.e., \mbox{$\kappa \leqslant \tilde{\kappa}$}, because $\PD$ depends only on the kurtosis $\kappa$ of the constellation. This reduces the computational complexity and leads to the following stochastic optimization problem
\begin{align}
    \max_{\mathcal{X}, P(X)} \ & \sum_{m=1}^{M}I(\RV{b}_m;\RV{L}_m) \\
    \text{s.t. } & \kappa \leqslant \tilde{\kappa} \tag{C\num{1}} \label{eq:system_model:C1} \\
    & \mathbb{E}_{\RV{X}} \{ | \RV{X} |^2 \} = 1 \tag{C\num{2}} \label{eq:system_model:C2} \\
    & \textstyle \! \sum_{X}P(X) = 1 \tag{C\num{3}} \label{eq:system_model:C3} \\
    & P(X) \geqslant 0, \quad \forall X \in \mathcal{X}. \tag{C\num{4}} \label{eq:system_model:C4}
\end{align}

Here, constraint~(\ref{eq:system_model:C2}) ensures that the constellation has unit power, while the constraints~(\ref{eq:system_model:C3}) and~(\ref{eq:system_model:C4}) enforce that $P(x)$ satisfies the properties of a probability mass function.

\section{Constellation Shaping using Autoencoders}
In this section, we propose a bitwise \ac{AE} framework that incorporates the sensing constraint~(\ref{eq:system_model:C1}), enabling the joint optimization of both geometric and probabilistic shaping for \ac{SC}, as illustrated in Fig.~\ref{fig:AE:setup}. The \ac{AE} concept enables end-to-end numerical optimization of communication systems. In our setup, the receiver is a Gaussian demapper and only the constellation of the transmitter is trainable. Depending on the shaping method (geometric, probabilistic, or joint), we optimize the constellation points, their probabilities, or both. These trainable parameters, shown as orange blocks in Fig.~\ref{fig:AE:setup}, are implemented as linear layers without bias. For more details on \ac{AE}-based constellation optimization, we like to refer the reader to previous works~\cite{aref_end--end_2022, stark_joint_2019, rode_end--end_2023}.

For a batch size $B$, the distribution sampler generates random indices $\vect{i} \in \{1, \ldots, \tilde{M} \}^B$ according to the input distribution \mbox{$P(x) \triangleq \vect{P}$}. To ease optimization, we use the Gumbel-softmax trick and optimize the unconstrained probabilities $\tilde{\vect{P}} \in \mathbb{R}^{\tilde{M}}$ which are passed through a softmax layer to ensure that $\vect{P}$ satisfies the probability distribution properties~(\ref{eq:system_model:C2}) and~(\ref{eq:system_model:C3})~\cite{stark_joint_2019}.

These indices are then mapped to bits~$\vect{b}$ and one-hot vectors~$\vect{i}_{\mathrm{oh}}$. To select the transmit symbols, the one-hot vectors are pointwise multiplied with the normalized constellation points $\vect{X}$. Similar to the probabilities, we optimize the unconstrained constellation points $\tilde{\vect{X}}$, which are passed through a power normalization layer to obtain the unit power constellation points \mbox{$\mathcal{X} \triangleq \vect{X}$}.

The selected constellation symbols are then transmitted over an \ac{AWGN} channel~(\ref{eq:system_model:comm_channel}) and the \acp{LLR}~(\ref{eq:system_model:LLR}) are computed for each bit using a classical Gaussian demapper~\cite{Ivanov_BICM}. Finally, the \ac{GMI}~(\ref{eq:system_model:GMI}), which should be maximized, is computed from the bits and \acp{LLR}.

We showed in Sec.~\ref{sec:ch2:system_model} that the detection probability constraint can be reformulated as a kurtosis constraint~(\ref{eq:system_model:C1}). Therefore, we propose a sensing loss term that depends on the kurtosis of the constellation $\kappa$ to satisfy the detection probability constraint
\begin{equation}
    \mathcal{L}_{\text{sens}} = {\begin{cases}
        0, &   \kappa \leqslant \tilde{\kappa} \\
        d (\kappa - \tilde{\kappa}) & \kappa > \tilde{\kappa}.
    \end{cases}}
    \label{eq:AE:sensing_loss}
\end{equation}
The penalty factor \mbox{$d \in \mathbb{R}^{+}$} controls the strength of the penalty if the kurtosis threshold $\tilde{\kappa}$ is violated. To maximize the integration gain, the overall loss function combines both \ac{SC} performance
\begin{equation}
    \mathcal{L} = \underbrace{\frac{M - \mathrm{GMI}}{M}}_{\text{Communications}} + \mathcal{L}_{\text{sens}}.
    \label{eq:AE:overall_loss}
\end{equation}
Note that this loss function does not strictly enforce the sensing constraint. However, the communications loss term is normalized between $[0,1]$ and by selecting the penalty factor $d$ sufficiently large, the sensing loss term dominates if the kurtosis constraint is violated, effectively enforcing the sensing constraint.

\section{Simulation Results}
\label{sec:simulation}
In this section, we validate our derivation of the constellation-dependent detection probability through simulations and compare the three constellation shaping methods in terms of their \ac{SC} performance, as well as the resulting trade-off. Throughout this section, we optimize constellations with $M = \SI{6}{bit \per symbol}$ for each kurtosis constraint $\tilde{\kappa} \in [1,2]$ independently, under a communications \ac{SNR} of $\mathrm{SNR}_\text{c} = \SI{10}{dB}$ and a penalty factor $d = 3$. The constellations are initialized as \ac{QAM} and optimized by minimizing the loss function~(\ref{eq:AE:overall_loss}) using the Adam optimizer. During training, we increase the batch size increases from 500 to 10,000, while we decrease the learning rate, with the initial value depending on the shaping method. As discussed in Sec.~\ref{sec:ch2:system_model}, the strongest sensing constraint $\tilde{\kappa} = 1$ should maximize sensing performance. On the contrary, a larger kurtosis constraint $\tilde{\kappa}$ is expected to improve communications performance, which is maximized for a circular symmetric Gaussian, that has a kurtosis of $\kappa = \num{2}$. A kurtosis constraint $\tilde{\kappa}$ between these two extremes should yield a trade-off between \ac{SC} performance.

\begin{figure}[!t]
    \vspace{0.8cm}
    \input{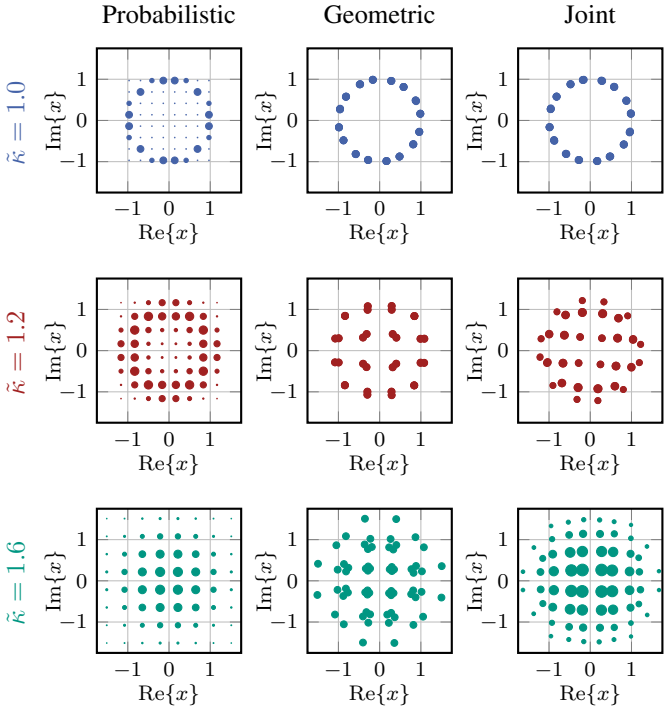}
    \vspace{-0.8cm}
    \caption{Optimized constellations for geometric, probabilistic and joint constellation shaping under various kurtosis constraints $\tilde{\kappa}$. The size of each constellation point is proportional to its probability and each constellation point has an associated bit label, which is omitted for clarity.}
    \label{fig:res:constellations}
    \vspace{-0.3cm}
\end{figure}

\subsection{Optimized Constellations}
Fig.~\ref{fig:res:constellations} shows the optimized constellations using geometric, probabilistic and joint constellation shaping for three kurtosis constraints $\tilde{\kappa}$. We note that the strongest sensing constraint $\tilde{\kappa} = \num{1.0}$ results in unit modulus constellations to reduce the sensing loss term~(\ref{eq:AE:sensing_loss}). In this case, both geometrically and jointly shaped constellations resemble a \ac{PSK} with overlapping constellation points. For probabilistic constellation shaping, a \ac{PSK} is only approximated because the constellation points which have the same power do not have equal distances. For a loose sensing constraint, the constellations approximate a Gaussian distribution to maximize communications performance. Between these two extremes (\mbox{$\tilde{\kappa} = \num{1.2}$}), a balance between Gaussian and unit modulus distribution is learned depending on the kurtosis constraint $\tilde{\kappa}$. Interestingly, the jointly shaped constellations exhibit more geometric shaping characteristics for a small kurtosis constraint $\tilde{\kappa}$ and more probabilistic behavior as the kurtosis constraint $\tilde{\kappa}$ increases.

\subsection{Sensing Performance}
\begin{figure}
    \centering
    \input{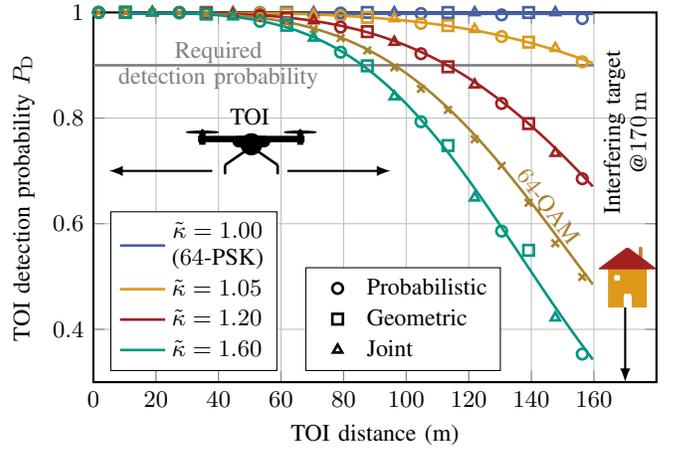}
    \vspace{-0.8cm}
    \caption{Derived and simulated detection probability of a \ac{TOI}, e.g. a drone, with an \ac{RCS} of \mbox{$\sigma_{\mathrm{RCS}} = \SI{0.1}{\metre\squared}$} following a Swerling-1 model in the presence of an interfering target, e.g. a building, at \SI{170}{\metre} with an \ac{RCS} of \mbox{$\sigma_{\mathrm{RCS}} = \SI{500} {\metre\squared}$} following a Swerling-0 model~\cite{richards_fundamentals_2014}.}
    \label{fig:res:sensing_performance}
    \vspace{-0.4cm}
\end{figure}

In Fig.~\ref{fig:res:sensing_performance}, we consider a scenario with two targets: a distant interfering target and a nearby \ac{TOI}, which should be detected and whose distance is varied. Multi-target scenarios are of particular interest because the average \ac{SINR}~(\ref{eq:system_model:SNR_TOI}) depends on the kurtosis and the power reflected by all targets. The simulation setup follows~\cite{braun_ofdm_2014}, with the simulation parameters being the FR2 case from~\cite{mandelli_survey_2023} with only one \ac{OFDM} symbol. For the \ac{CA}-\ac{CFAR}, we assume a false alarm rate of \mbox{$\PFA = 10^{-3}$} and a sliding window length of $N=\num{100}$. 

We found that the simulated detection probabilities (markers) align well with the analytical constellation-dependent detection probabilities (curves), verifying our derivation in Sec.~\ref{sec:ch2:system_model}. The minor discrepancies stem from the fact that the kurtosis $\kappa$ of the optimized constellations does not exactly match the kurtosis constraint $\tilde{\kappa}$. As expected from (\ref{eq:system_model:detection_prob}) and (\ref{eq:system_model:SNR_TOI}), the detection probability decreases with increasing kurtosis $\kappa$ and depends only on the kurtosis of the constellation, irrespective of the specific shaping method. Moreover, our simulations demonstrate that constellation shaping enables a dynamic adjustment of the detection range. For example, for a required detection probability of $P_{\mathrm{D}} = \num{0.9}$, the detection range can be varied from $\SI{90}{\metre}$ to beyond $\SI{160}{\metre}$ by modifying the kurtosis $\kappa$ of the constellation.

\begin{figure}[!b]
    \centering
    \vspace{-0.3cm}
	\input{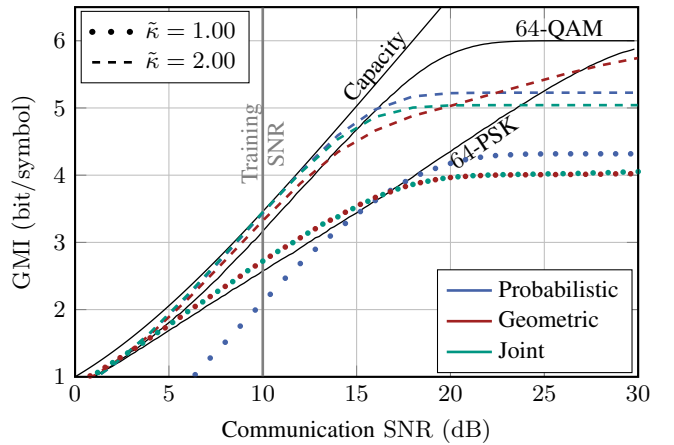}
    \vspace{-0.8cm}
    \caption{Simulated communications performance (\ac{AIR}) of the optimized constellations in comparison to legacy constellation formats.}
    \label{fig:res:comm}
\end{figure}

\subsection{Communications Performance}
Fig.~\ref{fig:res:comm} shows the \ac{GMI} as a function of the communications \ac{SNR} for various sensing constraints $\tilde{\kappa}$. For \mbox{$\tilde{\kappa} = 2$}, all shaping methods reduce the gap to capacity and outperform the conventional \num{64}-\ac{QAM} across an \ac{SNR} range of $\SI{10}{dB}$. For large \ac{SNR} values, the \ac{GMI} of the probabilistically and jointly shaped constellations plateaus below $\SI{6}{bit/symbol}$, since the constellations are optimized for a communications \ac{SNR} of \mbox{$\mathrm{SNR}_\text{c} = \SI{10}{dB}$} resulting in constellations with lower entropy and, consequently, a lower maximum \ac{GMI}. For \mbox{$\tilde{\kappa} = 1$}, the \ac{GMI} of the shaped constellations is similar to that of the \ac{PSK}, although probabilistic constellation shaping performs slightly worse due to unequal distances between the constellation points and the absence of Gray coding. Both geometric and joint constellations outperform the \num{64}-\ac{PSK} because their resulting constellations resemble a \num{16}-\ac{PSK} which achieves a higher \ac{GMI} at low \ac{SNR} values.

\begin{figure}
\centering
    \input{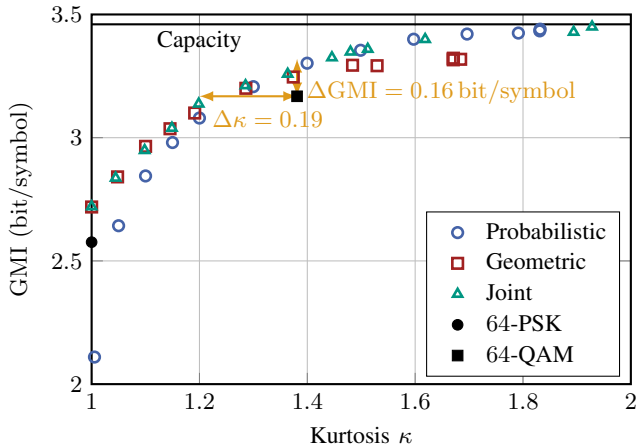}
    \vspace{-0.7cm}
    \caption{Comparison of the \ac{SC} trade-off between the shaping methods. Joint constellation achieves the performance of the superior shaping method for all kurtosis constraints.}
    \label{fig:results:random_determinisitic}
    \vspace{-0.3cm}
\end{figure}

\subsection{Communications-sensing Trade-off}
Finally, Fig.~\ref{fig:results:random_determinisitic} shows the \ac{GMI} as a function of the kurtosis $\kappa$ for the three constellation shaping methods, which effectively illustrates the communications-sensing (random-deterministic) trade-off. We note that a higher \ac{GMI} can be achieved given a larger kurtosis $\kappa$, with a particularly steep improvement at lower kurtosis values. Even without a kurtosis constraint $\tilde{\kappa}$, the kurtosis $\kappa$ of the optimized constellation saturates below $\kappa = \num{2}$ due to the low maximum amplitude of the optimized constellations compared to a Gaussian having infinitely extended tails.

We observe that probabilistic constellation shaping effectively approaches the capacity and outperforms geometric constellation shaping if the sensing constraint is loose. On the contrary, geometric constellation shaping outperforms probabilistic constellation shaping for strict sensing constraints $\tilde{\kappa} < 1.3$. This makes geometric and probabilistic constellation shaping well-suited for applications where sensing or communications performance is prioritized, respectively.

\textbf{Remark:} Joint constellation shaping combines the strengths of both approaches, maximizing the integration gain. It performs similarly to geometric and probabilistic constellation shaping at low and high kurtosis values, respectively. Compared to a conventional \num{64}-\ac{QAM}, joint constellation shaping increases the \ac{GMI} by \SI{0.16}{bit/symbol} and reduces the kurtosis $\kappa$ by \num{0.19} while maintaining an equivalent \ac{SC} performance. Additionally, while \num{64}-\ac{PSK} and \num{64}-\ac{QAM} offer only two discrete operating points, joint constellation shaping allows for continuous adjustment between \ac{SC} performance, providing greater flexibility. This makes joint constellation shaping a promising candidate for future 6G mobile communications systems, where maximizing performance and delivering a flexible trade-off between \ac{SC} capabilities will be crucial.

\section{Conclusion}
In this work, we employed an \ac{AE} framework to optimize geometric, probabilistic, and joint constellation shaping to improve \ac{SC} performance simultaneously. We derived that the detection probability depends only on the kurtosis of the constellation and is therefore independent of the applied shaping technique. Our simulations demonstrate that geometric shaping achieves a higher \ac{GMI} under strict sensing constraints, while probabilistic shaping performs better under relaxed sensing constraints. Notably, our proposed joint shaping approach combines the strengths of both geometric and probabilistic constellation shaping, significantly outperforming legacy constellation formats. This makes joint constellation shaping a promising candidate for 6G \ac{ISAC} systems, which need to deliver \ac{SC} capabilities efficiently and dynamically.


\begin{thebibliography}{10}
	\providecommand{\url}[1]{#1}
	\csname url@samestyle\endcsname
	\providecommand{\newblock}{\relax}
	\providecommand{\bibinfo}[2]{#2}
	\providecommand{\BIBentrySTDinterwordspacing}{\spaceskip=0pt\relax}
	\providecommand{\BIBentryALTinterwordstretchfactor}{4}
	\providecommand{\BIBentryALTinterwordspacing}{\spaceskip=\fontdimen2\font plus
		\BIBentryALTinterwordstretchfactor\fontdimen3\font minus
		\fontdimen4\font\relax}
	\providecommand{\BIBforeignlanguage}[2]{{%
			\expandafter\ifx\csname l@#1\endcsname\relax
			\typeout{** WARNING: IEEEtran.bst: No hyphenation pattern has been}%
			\typeout{** loaded for the language `#1'. Using the pattern for}%
			\typeout{** the default language instead.}%
			\else
			\language=\csname l@#1\endcsname
			\fi
			#2}}
	\providecommand{\BIBdecl}{\relax}
	\BIBdecl
	
	\bibitem{wild_6g_2023}
	T.~Wild, A.~Grudnitsky, S.~Mandelli, M.~Henninger, J.~Guan, and F.~Schaich,
	``{6G} integrated sensing and communication: From vision to realization,'' in
	\emph{Proc. European Radar Conference {(EuRAD)}}, Berlin, Germany, Sep. 2023,
	pp. 355--358.
	
	\bibitem{liu_integrated_2022}
	F.~Liu, Y.~Cui, C.~Masouros, J.~Xu, T.~X. Han, Y.~C. Eldar, and S.~Buzzi,
	``Integrated sensing and communications: Toward dual-functional wireless
	networks for {6G} and beyond,'' \emph{{IEEE} J. Sel. Areas Commun.}, vol.~40,
	no.~6, pp. 1728--1767, Jun. 2022.
	
	\bibitem{Lu_challenges}
	S.~Lu \emph{et~al.}, ``Integrated sensing and communications: Recent advances
	and ten open challenges,'' \emph{{IEEE} Internet Things J.}, vol.~11, no.~11,
	pp. 19\,094--19\,120, Jun. 2024.
	
	\bibitem{keskin_fundamental_2024}
	M.~F. Keskin, M.~M. Mojahedian, C.~Marcus, O.~Eriksson, A.~Giorgetti,
	J.~Widmer, and H.~Wymeersch, ``Fundamental trade-offs in monostatic {ISAC}: A
	holistic investigation towards {6G},'' Aug. 2024, preprint, available at
	\url{https://arxiv.org/abs/2401.18011v2}.
	
	\bibitem{xiong_torch_2024}
	Y.~Xiong, F.~Liu, K.~Wan, W.~Yuan, Y.~Cui, and G.~Caire, ``From torch to
	projector: Fundamental tradeoff of integrated sensing and communications,''
	\emph{IEEE BITS the Information Theory Magazine}, 2024.
	
	\bibitem{stark_joint_2019}
	M.~Stark, F.~Ait~Aoudia, and J.~Hoydis, ``Joint learning of geometric and
	probabilistic constellation shaping,'' in \emph{Proc. {IEEE} {GLOBECOM}
		{Workshops}}, Waikoloa, HI, USA, Dec. 2019.
	
	\bibitem{aref_end--end_2022}
	V.~Aref and M.~Chagnon, ``End-to-end learning of joint geometric and
	probabilistic constellation shaping,'' in \emph{Proc. Opt. Fiber Commun.
		Conf. {(OFC)}}, San Diego, CA, USA, Mar. 2022, p. W4I.3.
	
	\bibitem{du_reshaping_2023}
	Z.~Du, F.~Liu, Y.~Xiong, T.~X. Han, Y.~C. Eldar, and S.~Jin, ``Reshaping the
	{ISAC} tradeoff under {OFDM} signaling: A probabilistic constellation shaping
	approach,'' \emph{{IEEE} Trans. Signal Process.}, 2024.
	
	\bibitem{Yang_Constellation_Design}
	X.~Yang, R.~Zhang, D.~Zhai, F.~Liu, R.~Du, and T.~X. Han, ``Constellation
	design for integrated sensing and communication with random waveforms,''
	\emph{{IEEE} Wireless Commun. Lett.}, 2024.
	
	\bibitem{richards_fundamentals_2014}
	M.~A. Richards, \emph{\BIBforeignlanguage{en}{Fundamentals of {Radar} {Signal}
			{Processing}}}, 2nd~ed.\hskip 1em plus 0.5em minus 0.4em\relax New York:
	McGraw-Hill Education, 2014.
	
	\bibitem{Ivanov_BICM}
	M.~Ivanov, C.~Häger, F.~Brännström, A.~Graell~i Amat, A.~Alvarado, and
	E.~Agrell, ``On the information loss of the max-log approximation in {BICM}
	systems,'' \emph{{IEEE} Trans. Inf. Theory}, vol.~62, no.~6, pp. 3011--3025,
	Jun. 2016.
	
	\bibitem{liu_ofdm_2024}
	F.~Liu, Y.~Zhang, Y.~Xiong, S.~Li, W.~Yuan, F.~Gao, S.~Jin, and G.~Caire,
	``{OFDM} achieves the lowest ranging sidelobe under random {ISAC}
	signaling,'' Jul. 2024, preprint, available at
	\url{https://arxiv.org/abs/407.06691}.
	
	\bibitem{rode_end--end_2023}
	A.~Rode, B.~Geiger, S.~Chimmalgi, and L.~Schmalen, ``End-to-end optimization of
	constellation shaping for {Wiener} phase noise channels with a differentiable
	blind phase search,'' \emph{J. Lightw. Technol.}, vol.~41, no.~12, pp.
	3849--3859, Jun. 2023.
	
	\bibitem{braun_ofdm_2014}
	M.~Braun, ``{OFDM} radar algorithms in mobile communication networks,'' Ph.D.
	dissertation, Karlsruhe Institute of Technology, 2014.
	
	\bibitem{mandelli_survey_2023}
	S.~Mandelli, M.~Henninger, M.~Bauhofer, and T.~Wild, ``Survey on integrated
	sensing and communication performance modeling and use cases feasibility,''
	in \emph{Proc. {International} {Conference} on {6G} {Networking} ({6GNet})},
	Paris, France, Oct. 2023.
	
\end{thebibliography}
\end{document}